\documentclass[twocolumn]{aastex631}

\usepackage[T1]{fontenc}
\usepackage{ae,aecompl}

\usepackage{graphicx}	
\usepackage{amsmath}	
\usepackage{amssymb}	
\usepackage{soul}
\usepackage{subfigure}

\newcommand{\citelink}[1]{\hyperlink{cite.\{biblio.bib}}
\newcommand{\ssymbol}[1]{^{\@fnsymbol{#1}}}

\defcitealias{IPTA}{IPTA DR2}
\defcitealias{Ming18}{M18}
\defcitealias{NE2001}{NE2001}
\defcitealias{YMW16}{YMW16}
\defcitealias{Dller08}{D08}
\defcitealias{Desvignes}{D16}
\defcitealias{Nano}{A18}
\defcitealias{Ding}{D20}
\defcitealias{Verbiest}{V08}
\defcitealias{Nicastro}{N95}
\defcitealias{Bassa}{B16}
\defcitealias{Kaplan}{K16}
\defcitealias{Rear16}{R16}
\defcitealias{Durant}{D12}
\defcitealias{Sanchez}{S20}
\defcitealias{J1910_discovery}{S05}
\defcitealias{Gonzalez11}{G11}
\defcitealias{Ding23}{D23}
\defcitealias{Deneva}{De12}

\begin{document}{}

\title{Improving Distances to Binary Millisecond Pulsars with \emph{Gaia}}

\author[0000-0002-6437-5229]{Abigail Moran}
\affiliation{Department of Physics, University of Connecticut, 196 Auditorium Road, U-3046, Storrs, CT 06269-3046, USA}

\author[0000-0002-4307-1322]{Chiara M. F. Mingarelli}
\affiliation{Department of Physics, Yale University, New Haven, CT, 06520, USA}
\affiliation{Department of Physics, University of Connecticut, 196 Auditorium Road, U-3046, Storrs, CT 06269-3046, USA}
\affiliation{Center for Computational Astrophysics, Flatiron Institute, 162 Fifth Ave, New York, NY, 10010, USA}

\author[0000-0001-9907-7742]{Megan Bedell}
\affiliation{Center for Computational Astrophysics, Flatiron Institute, 162 Fifth Ave, New York, NY, 10010, USA}

\author[0000-0003-1884-348X]{Deborah Good}
\affiliation{Department of Physics, University of Connecticut, 196 Auditorium Road, U-3046, Storrs, CT 06269-3046, USA}

\author[0000-0002-5151-0006]{David N. Spergel}
\affiliation{Center for Computational Astrophysics, Flatiron Institute, 162 Fifth Ave, New York, NY, 10010, USA}

\begin{abstract}
Pulsar distances are notoriously difficult to measure, and play an important role in many fundamental physics experiments, such as pulsar timing arrays (PTAs). Here we perform a cross-match between International PTA pulsars (IPTA) and \emph{Gaia}'s DR2 and DR3. We then combine the IPTA pulsar's parallax with its binary companion's parallax, found in \emph{Gaia}, to improve the distance measurement to the binary. We find 7 cross-matched IPTA pulsars in \emph{Gaia} DR2, and when using \emph{Gaia} DR3, we find 6 IPTA pulsar cross-matches, but with 7 \emph{Gaia} objects. Moving from \emph{Gaia} DR2 to \emph{Gaia} DR3, we find that the \emph{Gaia} parallaxes for the successfully cross-matched pulsars improved by $53\%$, and pulsar distances improved by $29\%$. Finally, we find that binary companions with a $<3.0\sigma$ detection are unreliable associations, setting a high bar for successful cross-matches.

\end{abstract}

\keywords{stars: distances, pulsars: general, gravitational waves, binary systems}

\section{Introduction} \label{sec:intro}

Millisecond pulsars (MSPs) are valuable probes in several areas of astrophysics, including gravitational wave detection with pulsar timing arrays (PTAs; \citealt{sazhin,det79,HD83}), tests of general relativity (e.g. \citealt{GenRelTest}), and dark matter density mapping \citep{Philips}. First discovered in \cite{backer}, many MSPs have microsecond timing precision, making them some of the best clocks in nature. 

For these experiments, outcomes are improved by better pulsar distance measurements. {Measuring} these distances can be difficult since many MSPs are at kiloparsec distances, making precise parallax measurements difficult. For well timed pulsars (those with timing precision of at least $1 \mu s$ for pulsars $\sim 1$ kpc away; \citealt{Toscano}) timing parallax can be used to determine distances \citep{BackerHellings}. Similarly, for pulsars which have been imaged to sub-milliarcsecond precision, very long baseline interferometry (VLBI) can be used to measure distances \citep{Salter}. For some pulsars in binaries, we can estimate their distances by using the so-called kinematic distance measurement ($D_{k}$; \citealt{Shk70, Bell}). 

The distance to a pulsar can also be constrained indirectly -- the Dispersion Measure (DM) of a pulsar is the delay in the arrival of a pulse as a result of its travel through the ionized interstellar medium. These measurements therefore are dependant on the column density of electrons in the model of the galaxy. The two principal models currently in use are \cite{NE2001} and \cite{YMW16}, hereafter referred to as \citetalias{NE2001} and \citetalias{YMW16} respectively. However there is a great deal of uncertainty in these DM-based distance estimates, conservatively $20-40\%$ \citep{NE2001, YMW16}.

Binary pulsar companions thus offer a complementary path to making a distance measurement by finding an optical counterpart in e.g. \emph{Gaia}.
While \cite{Jennings} carried out a cross-match between pulsars with known companions and \emph{Gaia} DR2,{ here, as in \cite{Ming18}, hereafter \citetalias{Ming18}}, we combine both pulsar-timing-based parallaxes with \emph{Gaia}-based parallaxes to improve the overall distance measurement.
{This work includes and supercedes the results of \citetalias{Ming18}, carrying out and reporting the results of cross-matches with \emph{Gaia} DR2 and DR3.}
We improve on our cross-match statistics by looking at the temperature of the companion in addition to a novel False Alarm Probability (FAP) calculation.

{Compared with \emph{Gaia} DR2, DR3 provides measurements of parallax and proper motion for $10\%$ more objects \citep{Arenou}.}
{In addition to the millions of new sources, \emph{Gaia} DR3 also includes updated measurements of proper motion, sky position, parallax, and photometric parameters for over $96\%$ of DR2 sources. }

{The paper is laid out as follows: in \autoref{sec:binary_candidates} we describe how we identify pulsar companions in \emph{Gaia}, and compute the FAPs of these associations using two approaches: one is a chance association, and the other verifies the match via computing the temperature of the companion and cross-validating it. We report our results in \autoref{sec:results}, and close with a summary of our results and discussion of them in  \autoref{sec:discussion}.} 

All of the data analysis software which was used in this work is publicly available on github, written in Python at https://github.com/abby-moran/gaiaDR3-pulsars.

\section{Identifying Binary Candidates in \emph{Gaia}} \label{sec:binary_candidates}
\subsection{Cross-Matching IPTA DR2 with \emph{Gaia}}
We cross referenced the sky positions of MSPs in \cite{IPTA}, hereafter \citetalias{IPTA}, with objects in \emph{Gaia} DR2 and and DR3 \citep{GaiaDR2, gaiaDR3}. 
Using the proper motion and coordinates of each candidate match, we updated the \emph{Gaia} objects to the \citetalias{IPTA} epoch. We require that the object's position is within 3$\sigma$ of the pulsar's position in \citetalias{IPTA}. {We also require that the object has a parallax measurement, and that the proper motion of the object in RA and DEC are within $3 \sigma$ of the pulsar's \citetalias{IPTA} values.}
%

\subsection{False Alarm Probabilities: Chance Association?}
\label{subsec:FAP}
It is possible that in certain more crowded parts of the galaxy, pulsar systems have higher FAPs --- that is to say that the \emph{Gaia} object and the pulsar are in chance alignment rather than a binary system. 
In order to test the null hypothesis we calculate the FAP for each potential system. We do this by randomizing the pulsar's sky coordinates within three arcseconds in both right ascension and declination. This has {the added benefit} of taking into account the more crowded parts of the sky where the pulsars are more likely to randomly align with a \emph{Gaia} object. We then search around the pulsar within a radius of 3 arcseconds for \emph{Gaia} objects and repeat the process at least ${10^7}$ times. The number of trials in which a \emph{Gaia} object is found divided by the number of total trials gives the FAP. 
{We carry out this test using astrometric data from the relevant \emph{Gaia} data release; for associations detected in both DR2 and DR3 we report a FAP for each data release.} 

{We set our detection threshold to be $3.0 \sigma$, since in a cross-match between PSR J1949+3106 (hereafter J1949+3106) and \emph{Gaia} DR2, we found a $3.0 \sigma$ detection of a binary companion which was not found in \emph{Gaia} DR3.} We therefore only claim valid cross-matches for pulsars with $>{3.0}\sigma$ detections --- see \autoref{subsec:J1929} for more details. Systems with higher FAPs indicate tentative matches in need of further verification.

\subsection{False Alarm Probabilities: Temperatures} \label{subsec:temps}
Many IPTA pulsars have known binary companions, and some of these companions have published effective temperatures. In an effort to further validate our \emph{Gaia}-IPTA cross-matches, we calculate the temperature of each cross-matched \emph{Gaia} object based on \emph{Gaia} DR3 photometric data.
We find the magnitude for the blue passband, $G_\mathrm{BP}$, and for red light, $G_\mathrm{RP}$  for each source. We calculate the effective temperature following as in \cite{Jordi}: 
\begin{equation}
\!\!\log (T_\mathrm{eff}\!) \!=\! 3.999\!-\!0.654(\!C_\mathrm{XP}\!)\!+\!0.709(\!C_\mathrm{XP}\!)^2\!-\!0.316(\!C_\mathrm{XP}\!)^3 \!\!\!
\label{eq:temp}
\end{equation}
where $C_\mathrm{XP}$ is given by $G_{\mathrm{BP}}-G_{\mathrm{RP}}$. For objects with $C_\mathrm{XP} < 1.5$ \autoref{eq:temp} has a standard error of $\sigma_{T} = 0.0046T_\mathrm{eff}$. The error introduced by this model increases as we approach $C_\mathrm{XP} = 1.5$, which is the case for several of our cross-matches.
While \emph{Gaia} does not report individual uncertainties on the magnitudes used in \autoref{eq:temp}, {\cite{gaiaDR3} estimates that for stars in \emph{Gaia} DR3 with $G$ $\approx$ 20, the $G_{\mathrm{BP}}$ error is 180 mmag, and 52 mmag for $G_\mathrm{{RP}}$.} Our final error is the quadrature sum of this uncertainty, and the uncertainty introduced by \autoref{eq:temp}.

For objects with $C_\mathrm{XP} >1.5$, we explore an alternate route to estimate the T\textsubscript{eff}. We compare the photometric data of the \emph{Gaia} DR3 object to the synthetic catalog in \cite{Jordi}: based on the object's magnitude in blue (G\textsubscript{BP}) and red (G\textsubscript{RP}) as well as its color (G\textsubscript{BP}-G\textsubscript{RP}), we find the synthetic star which is most similar to our \emph{Gaia} object. We then use the T\textsubscript{eff} of this similar object as an estimate for the companion's temperature.

\section{Results} \label{sec:results}

\begin{figure}[hb!]
\centering
\includegraphics[]{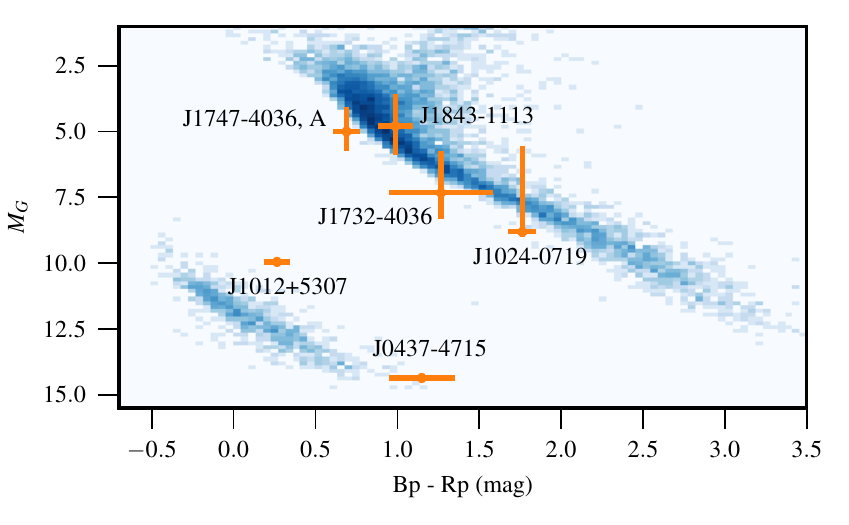}
	\caption {A color-magnitude plot displaying 6 of the companions to pulsars from this cross-match (orange) against a background of well-measured stars in the \emph{Gaia} catalog.}
    \label{fig:hrd}
\end{figure}

\begin{table*}[]
\centering
\begin{tabular} 
{|p{2.1cm}|p{1.7cm}|p{1.9cm}|p{1.5cm}|p{2.1cm}|p{2.85cm}|c|c|}
 \hline
 \hline
 Pulsar & RA & DEC & $P_b$ (d) & Companion Type & Reference & DR2 & DR3 \\
 \hline
 \hline
 J0437-4715 & 04:37:15.91 & $-$47:15:09.21 & 5.741 & White Dwarf & \citetalias{Verbiest}, \citetalias{Dller08}, \citetalias{Desvignes} & X & X \\
 J1012+5703&  10:12:33.44 & +53:07:02.30 & 0.6046 & White Dwarf & \citetalias{Nicastro}, \citetalias{Desvignes}, \citetalias{Nano}, \citetalias{Ding} & X & X \\ 
J1024-0719 &  10:24:38.68 & $-$07:19:19.43 &  $\sim10^4$ & Main sequence & \citetalias{Kaplan}, \citetalias{Bassa} & X & X \\
$^\dagger$J1732-5049 & 17:32:47.77 & $-$50:49:00.21 & 5.263 & -- & \citetalias{Rear16}& X & X \\
$^\dagger$J1747-4036 &  17:47:48.72 & $-$40:36:54.78 & -- & -- & -- & -- & X\\
{$^\dagger$J1843-1113} & 18:43:41.26 & $-$11:13:31.07 & -- & -- & -- & X & X\\
{J1910+1256} & 19:10:09.70 & +12:56:25.49 & 58.47 & -- &  \citetalias{J1910_discovery}, \citetalias{Desvignes}, \citetalias{Ding23} & X & --\\
\textsuperscript{*}{J1949+3106} & 19:49:29.64 & $+$31:06:03.80 & 1.950 & White Dwarf & \citetalias{Deneva}& X & --\\
\hline
\hline
 \end{tabular}
 \caption{{A summary of the pulsars for which companions were identified in \emph{Gaia} DR2 and DR3 with previously published data. Improved distances {for those detected in \emph{Gaia} DR3} are reported in Table \ref{Tab:res}. Pulsar positions are from \protect\citetalias{IPTA}, while binary periods and the companion types when known are from references cited in the `Reference' column. Unknown companion types are denoted by `--'. 
 In the last two columns, an `X' indicates a positive cross-match was found (see Sec. \ref{sec:binary_candidates}) in the given \emph{Gaia} data release and a `-' indicates no companion was identified. A `$^\dagger$' indicates a weak ($< 3.0\sigma$) \emph{Gaia} association in both DR2 and DR3.
 *J1949+3106 had a companion identified in \emph{Gaia} DR2 but not in \emph{Gaia} DR3. We therefore believe this to be a false cross-match result from DR2, and explore the implications of this in \autoref{subsec:J1929}.}
\\
}
\label{Tab:background}
 \end{table*}

We have identified six {candidate} binary companions to IPTA pulsars in  \emph{Gaia} DR3: these are PSRs J0437-4715, J1012+5307, J1024-0719, J1732-5049, J1747-4036, {and J1843-1113} (see \autoref{Tab:background}) for more information). 
We report the FAPs of these detections and distance measurements based on \emph{Gaia} DR3 parallaxes combined with pulsar timing-based parallax measurements. Our results are also summarized in \autoref{Tab:res}.

We calculate the combined distance to the binary systems by multiplying the posteriors of PTA and/or VLBI-based parallax measurements with the \emph{Gaia} DR3 parallax value. We then apply the distance prior \cite{bailer-jones2021} and report the distance as the peak of the combined distance curve, with 16th and 84th percentiles as the distance errors. This distance prior includes a nonzero global parallax offset, though the impact of this on the final distance is negligible. {For comparison, we compute combined distances with \emph{Gaia} DR2 parallaxes using the \cite{BJ18} distance prior (\citetalias{Ming18}).}

{We do not to include kinematic distances when computing combined pulsar distance measurements, since proper motion errors are correlated between kinematic distance measurements and timing parallax.}

Notably, two candidate companions {in the \emph{Gaia} DR2 cross-match do not meet the requirements in DR3} — those associated with pulsars J1910+1256 and J1949+3106 (see \autoref{Tab:background}; \citetalias{Ming18}).
A match was found to PSR J1910+1256 in DR3, but was eliminated by our stringent requirements for proper motion and sky location accuracy.
We found no object associated with J1949+3106 in DR3 (see \autoref{subsec:J1929}) {despite finding a $3.0 \sigma$ association in \emph{Gaia} DR2. This is the basis of our detection threshold of $3.0 \sigma$. }


The pulsar companions we identified in \emph{Gaia} DR3 are shown on a color-magnitude plot in \autoref{fig:hrd}. Also on this diagram, in blue, we show a sample of well-measured stars from \cite{Gaiacollab}. {Magnitudes in \emph{Gaia} DR3 include E(B-V) dust corrections \citep{Riello}. Using our median distance values and a 3D dust map \citep{3D}, we find that this correction is zero for all sources in this study, except for the objects associated with J1024-0719 and J1843-1113, which have E(B-V)$ = 40$ and $718$ mmag, respectively.} 

{Below we discuss each binary pulsar system in \autoref{Tab:background}, except for PSR J1910+1256. Our results are structured as follows: first we discuss  J1949+3106, which we use to set our FAP threshold with. This is followed by weak \emph{Gaia} associations with PSRs J1732-5049, J1747-4036, and J1843-1113 (hereafter we omit the PSR prefix). We finish by describing detections of companion \emph{Gaia} objects to PSRs J0437-4715, J1012+5307, and J1024-0719 (dropping the PSR prefix from here).}


\begin{table*}
\centering
\begin{tabular} 
{|p{2.23cm}|p{1.3cm}|p{1.3cm}|p{1.9cm}|p{2cm}|p{1.9cm}|p{2.0cm}|p{2.4cm}|  }
 \hline
 \hline
 Pulsar & D\textsubscript{DM} (pc) \citetalias{NE2001} &D\textsubscript{DM} (pc) \citetalias{YMW16}& {Previous} parallax (mas) & \emph{Gaia} DR3 parallax (mas) & Combined parallax (mas) & Distance (pc) & Reference\\
 \hline
 \hline
J0437-4715   & 139    &156 &   6.37 $\pm$ 0.09 & 7.10 $\pm$ 0.52 & $6.40\pm 0.05$ & $156 _{-1.1}^{+1.1}$ & \citetalias{Dller08}, \citetalias{Desvignes}\\
 J1012+5703&   411  & 805  &0.92 $\pm$ 0.20 & 1.74 $\pm$ 0.29 & $1.17 \pm 0.02$ & $845^{+14}_{-14}$ & \citetalias{Desvignes}, \citetalias{Ding}, \citetalias{Ding23}\\
 J1024-0719   & 383 & 376 & 0.86 $\pm$ 0.15 & 0.86 $\pm$ 0.28 & $0.91 \pm 0.05$ & $1072^{+67}_{-49}$ & \citetalias{IPTA}, \citetalias{Ding23} \\
 $^\dagger$J1732-5049   & 1411 & 1875 & None & $-0.54 \pm 2.22$ &$-0.54 \pm 2.22$ & $3874^{+4100}_{-1400}$ & \citetalias{IPTA} \\
 $^\dagger$J1747-4036 [A]  & 3392 & 7152  & 0.4 $\pm$ 0.7 & $-0.88 \ \pm$ 0.46 &  $-0.49 \pm 0.38 $ & $6042^{+3200}_{-1700}$ & \citetalias{IPTA} \\
 $^\dagger$J1747-4036 [B]  & 3392 & 7152 & 0.4 $\pm$ 0.7 & 1.83 $\pm$ 0.97 & $0.87 \pm 0.55$ & $4028^{+3800}_{-1600}$& \citetalias{IPTA} \\
 $^\dagger$J1843-1113 & 1697 & 1705 & $0.69 \pm 0.33$ & $1.06 \pm 0.52$ & $0.80 \pm 0.28$ & $4568_{-1800}^{+3500}$ & \citetalias{Desvignes}\\
 \hline
 \hline
\end{tabular}
 \caption{Summary of results.
 {The previous parallax measurement is the parallax value from \citetalias{IPTA}. For PSRs J0437-4715 and J1012+5307 this is a VLBI measurement, and for PSRs J1024-0719, J1747-4036, and J1843-1113 this is a timing parallax. }
 The combined distances take into account the \emph{Gaia} DR3 parallax measurement and all other available parallax measurements. The asymmetric errors on these distances represent the 16th and 84th percentiles. Distances based on dispersion measures assume a standard error of $\pm20\%$ and are for comparison only.
{A  $^\dagger$ next to a pulsar's name indicates a weak ($<{3.0}\sigma$) \emph{Gaia} association which we hope to verify in future data releases.}
\\
}
\label{Tab:res}
\end{table*} 



\subsection{J1949+3106: False Association} \label{subsec:J1929}

\begin{figure*}
\centering
 	\subfigure[Right Ascension of J1949+3106's former \emph{Gaia} match]{\includegraphics[]{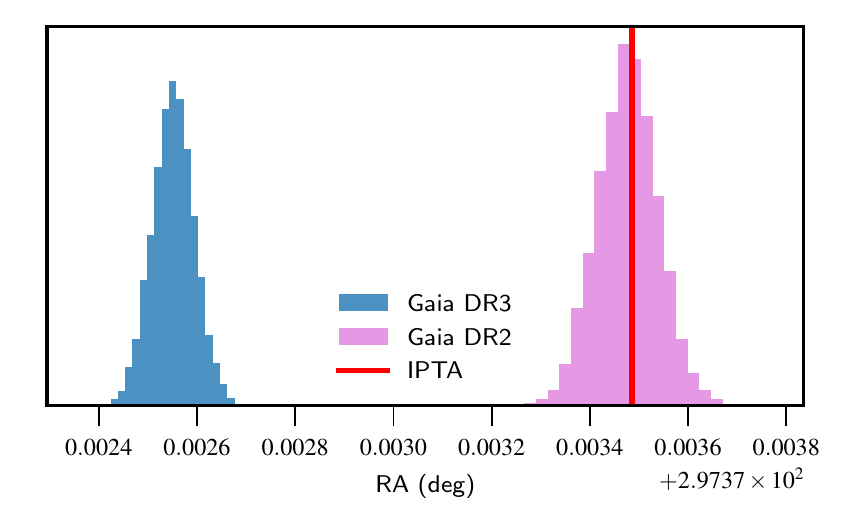}}
	\subfigure[Declination of J1949+3106's former \emph{Gaia} match]{\includegraphics[]{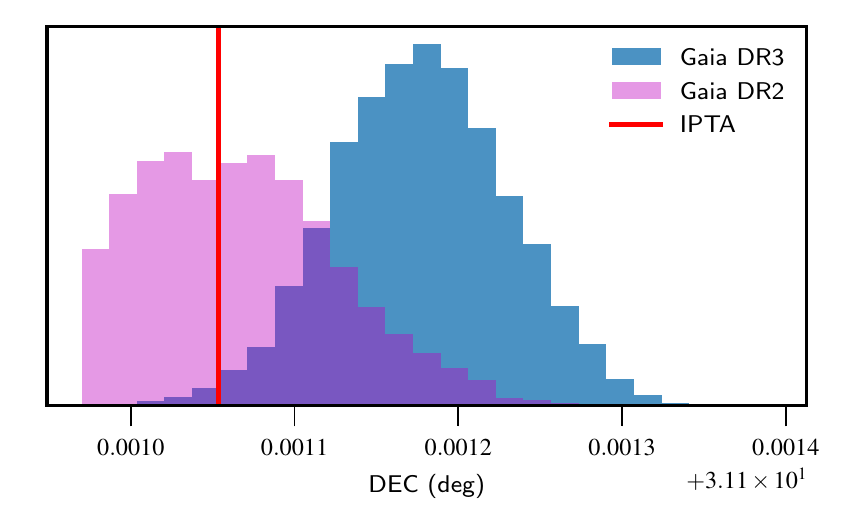}}
\caption {False association: the sky position of J1949+3106 (red IPTA bar) as compared to the object identified in \emph{Gaia} DR2 as the pulsar's companion (pink). The \emph{Gaia} DR3 data (blue) illustrate that this was a false association. {The \emph{Gaia} object's position differs from the pulsar's by $22\sigma$ in RA and $2.2\sigma$ in DEC.}}
\label{fig:J1949}
\end{figure*}

{In our \emph{Gaia} DR2 analysis we find a companion to J1949+3106 with a FAP of $3.19 \times 10^{-3}$, making this a ${3.0}\sigma$ detection.}
{However, in DR3, neither this DR2 candidate companion, \emph{Gaia} DR2 2033684263247409920, nor any other object meet the criteria to be a candidate for association with J1949+3106 (see \autoref{fig:J1949}). } 

While \citetalias{IPTA} confirms that a companion should exist, we find it unlikely that Gaia DR2 2033684263247409920 is this companion. 
{In \cite{Deneva} this companion is identified as a white dwarf with $M=0.85 \ M_{\odot}$. The dimmest objects in \emph{Gaia} DR3 have $G \sim 21$,  
and therefore this companion may be too dim for \emph{Gaia} to detect.}
In future data releases we will nonetheless continue to monitor J1949+3106.
We therefore require potential \emph{Gaia}companions to be detected at a $>{3.0}\sigma$ confidence to be called a confirmed companion.

\subsection{J1732-5049: {Weak Association}} \label{subsec:J1732}

J1732-5049 has a binary companion in \citelink{Perera}{IPTA DR2} with a period of 5.3 days. We tentatively identify this companion as \emph{Gaia} 5946288492263176704 (\citelink{Ming18}{M18}) in {DR2 and DR3 }. The parallax measurement to this object in DR2 is $-1.18 \pm 2.84$ and in DR3 is $-0.54 \pm 2.22$ mas. While negative parallaxes are unphysical, it may be useful to monitor this binary companion in the hopes of obtaining improved parallax measurements.

\begin{figure}[tbh]
\centering
\includegraphics[]{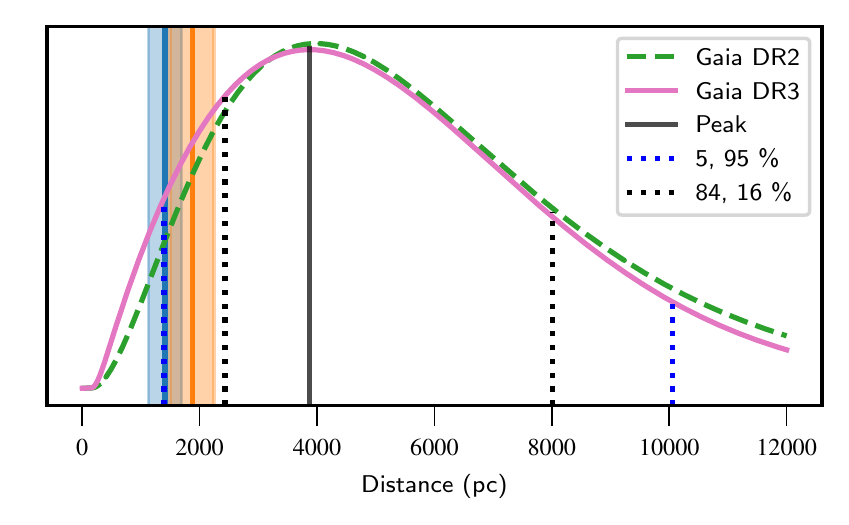}
	\caption {Distances to the system of pulsar J1732-5049 and its {candidate} binary companion. {These \emph{Gaia} parallaxes are the first potential measurements to the pulsar. Both parallax distances are more than double either DM distance (based on \citetalias{NE2001} in blue or \citetalias{YMW16} in yellow), which assume a standard error of $\pm 20\%$. This may indicate that these models have overestimated the electron density along this line-of-sight. }}
    \label{fig:J1732}
\end{figure}

We find the distance to this object is $3874^{+4100}_{-1400}$ pc using the DR3 parallax. When we report the 5th and 95th percentiles this is $3874^{+6200}_{-2500}$ pc (see \autoref{fig:J1732}). {Using the \emph{Gaia} DR2 parallax }the distance is is 3980 pc, with the 16th percentile as 4800 pc and the 84th as 11000 pc. 
Thus we see a $42\%$ decrease in the error. Since these distances are derived from negative parallaxes, the measurements are strongly dependent on the {distance priors put forth in \cite{BJ18} and \cite{bailer-jones2021} for DR2 and DR3, respectively. However, these priors make it possible to yield imprecise but nevertheless meaningful distance measurements from negative parallaxes.}

{
The DR3 identification has a FAP of $3.9 \times 10^{-3}$, making this a 2.9$\sigma$ association. Although this is an improvement from the $2.8 \sigma$ DR2 association,} we have learned that a ${3.0} \sigma$ detection is required for a reliable association, so we cannot yet claim that this is a detection. 

We now look to the temperature of the object to help us further understand the validity of the cross-match. With $G_\mathrm{BP}=21.40$ and $G_\mathrm{RP}=19.82$ this object has $C_\mathrm{XP} = 1.6$, thus we calculate the temperature using both \autoref{eq:temp} and via comparison to the synthetic catalog from \cite{Jordi}. Using the former, we calculate that this companion object has $T_\mathrm{eff}=3104 \pm {1204}$ K. However, the error on this is likely underestimated due to the constraints of the model. When the synthetic catalog method is used, we find $T_\mathrm{eff} \sim 4000 K$. {However, since there is no temperature for this companion in the literature, we are unable to further validate the match.}

\subsection{J1747-4036: {Weak Association}} \label{subsec:J1747}

\begin{figure*}
\centering
\setlength\fboxsep{0pt}
\setlength\fboxrule{0.25pt}
\setlength\fboxrule{0.25pt}
\subfigure[Right Ascension of Object A]{\includegraphics[scale=.9]{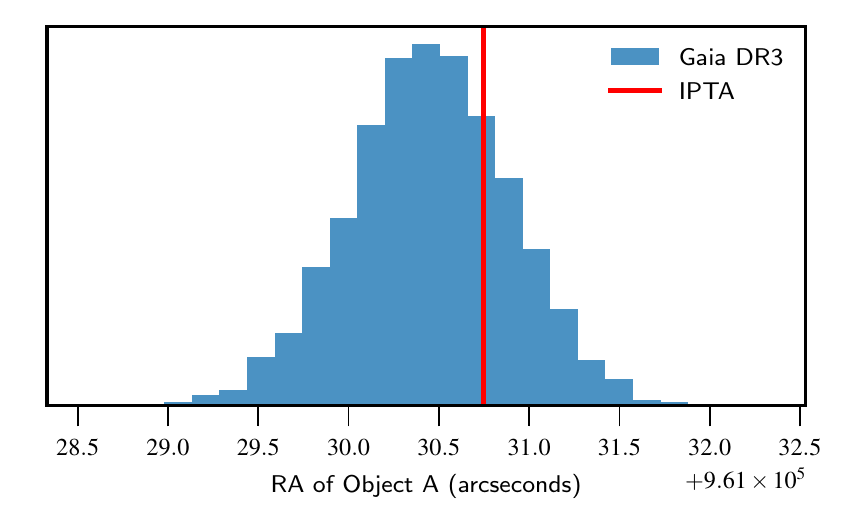}}
\subfigure[Declination of Object A]{\includegraphics[scale=.9]{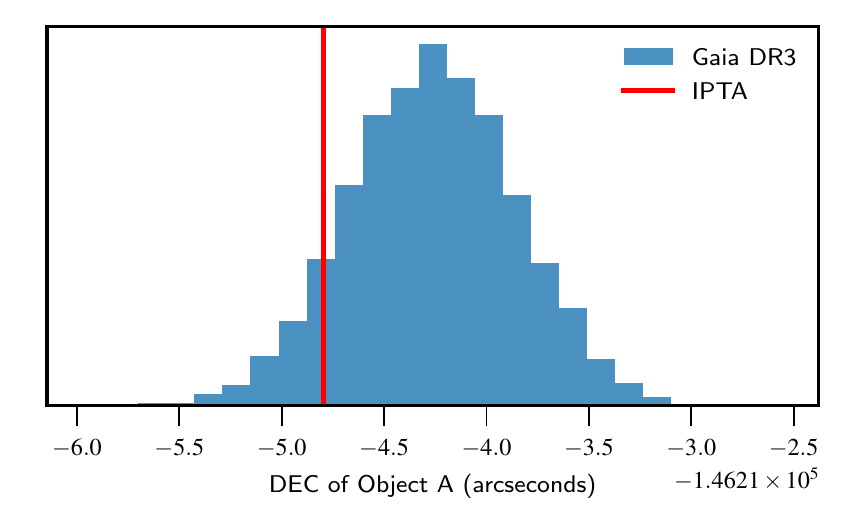}}
  \subfigure[Right Ascension of Object B]{\includegraphics[scale=.9]{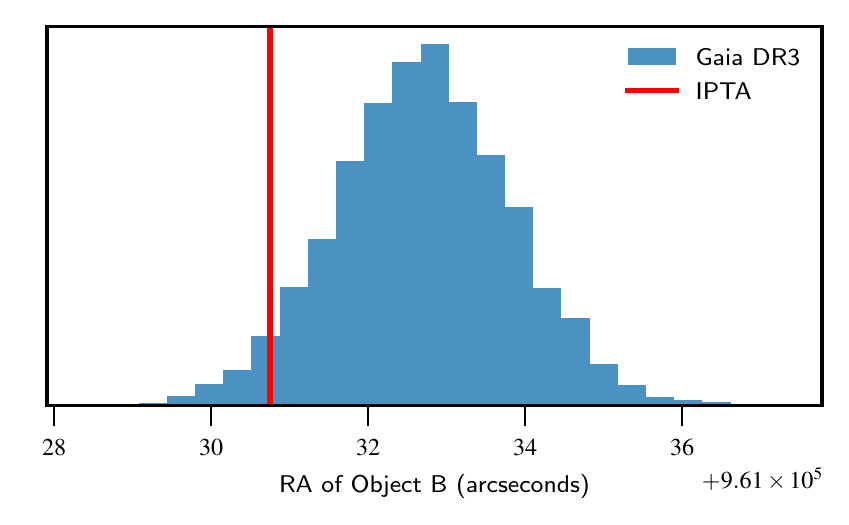}}
\subfigure[Declination of Object B]{\includegraphics[scale=.9]{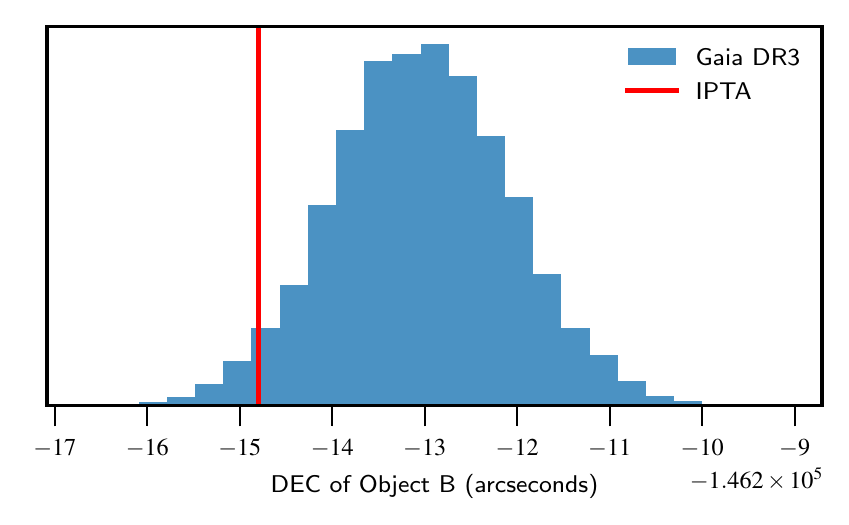}}
   \subfigure[Distances to J1747-4036]{\includegraphics[scale=.8]{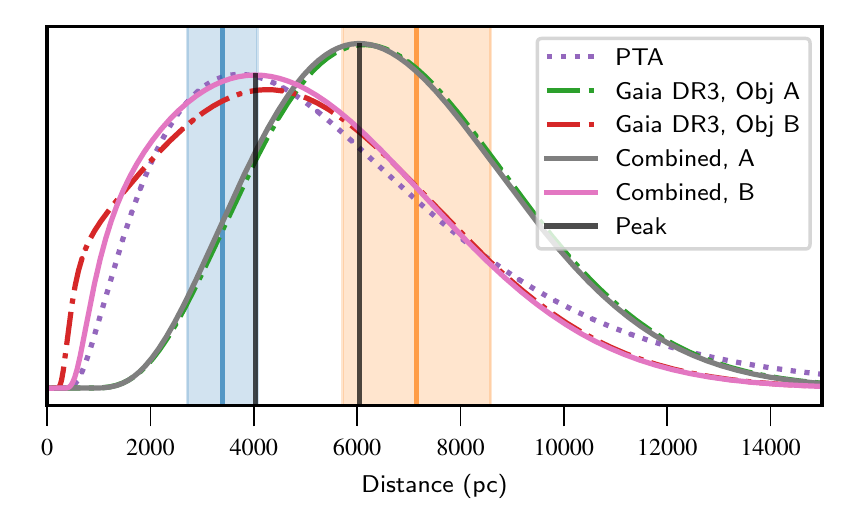}}
	\caption {Panels (a) through (d) show comparisons of the location of J1747-4036 (red) and candidate companion objects (blue) after updating the \emph{Gaia} object positions to the IPTA epoch. Both objects' positions are consistent with the IPTA position of J1747-4036. Panel (e) shows distances to the system of J1747-4036 in DR3. DM distances from \citetalias{NE2001} (blue strip) and \citetalias{YMW16} (yellow strip) are shown for illustrative purposes and are not included in the combined measurements.}
    \label{fig:J1747}
\end{figure*}

J1747-4036 has been classified as a solitary system \citep{Kerr}. Here, we identify two objects in \emph{Gaia} DR3 that are candidates for association with the MSP. These objects are \emph{Gaia} DR3 5957827763757710080 and 5957827763757708544, referred to hereafter as Object A and Object B, which are separated by 2.226 arcseconds. Either one of these objects may be a binary companion to J1747-4036, but with the current \emph{Gaia} data, we are unable to determine which, if not both, is the false match (see \autoref{fig:J1747}).

The \emph{Gaia} DR3 parallax measurements of Objects A and B are $-0.88 \pm 0.46$ and $1.83 \pm 0.97$ mas, respectively. These detections thus have S/Ns of $-1.91$ and $1.89$. 
In \citetalias{IPTA}, J1747-4036 has a parallax of $0.4 \pm 0.7$ mas. We combine this parallax measurement with the \emph{Gaia} parallax measurement of Object A and compute a distance of $6042^{+3200}_{-1700}$ pc.
When the IPTA value is instead combined with the parallax of Object B, the distance is $4028^{+3800}_{-1600}$ pc. 

J1747-4036's companion are co-located, and therefore have the same FAP. We find this FAP to be {$1.9 \times 10^{-2}$, making these weak $2.3\sigma$ associations. 
We therefore do not yet claim to have identified a new binary system, since the FAP is below the established threshold. }

{Although there is no known companion to this pulsar and thus no known temperature to compare our value with, we carry out our false alarm checks for completeness.} We find that Object A has $G_\mathrm{BP}=20.33$ and $G_\mathrm{RP}=19.14$. Using \autoref{eq:temp} we find that Object A has $T_\mathrm{eff}=4927 \pm {788}$ K, a temperature consistent with a cool white dwarf star \citep{Jordi}. There is no photometric data for Object B in \emph{Gaia} DR3.

The two \emph{Gaia} objects, A and B, have an angular separation of 2.226 arcseconds. This translates to a physical separation on the order of $10^3$ AU {using the \citetalias{IPTA} parallax distance}. Given the high degree of precision on \emph{Gaia} DR3 coordinates, it is thus improbable that these two objects are associated with one another. {It is more plausible that at least one object, and possibly both, are false match(es), particularly given that the associations are $< 3.0 \sigma$. Furthermore timing data for this pulsar point to an isolated system. This would therefore place a limit of $P_b \gtrapprox 1$ kyr on this potential binary \citep{binary_comp}.  So far, pulsar J1024-0719 appears to be the only pulsar we identify with a companion in an ultra-long orbit \citep{Bassa, Kaplan}.}
We are hopeful that future data releases will be able to verify or refute one or both matches. 

\subsection{J1843-1113: {Weak Association}} \label{subsec:J1843}

\begin{figure}[!htbp]
\centering
\includegraphics[]{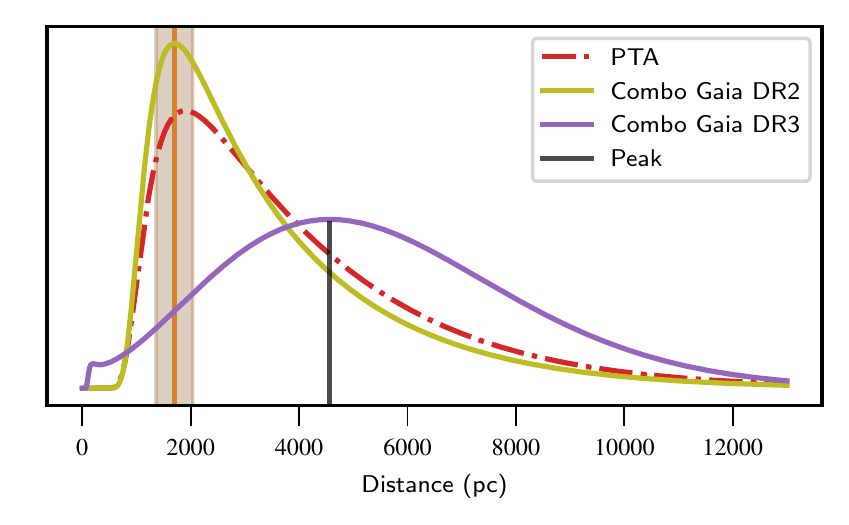}
\caption {{The combined parallax measurements to J1843-1113 and the resulting distances. Despite being a more precise parallax measurement, the \emph{Gaia} DR3 parallax corresponds to a less precise combined distance than when the \emph{Gaia} DR2 value is used as a result of its inconsistency with the PTA parallax measurement. The PTA measurements is also consistent with the \citetalias{NE2001} (blue strip) and \citetalias{YMW16} (yellow strip) DM distances. This casts further doubt on the association.}} 
\label{fig:J1843}
 \vspace{-1\baselineskip}
\end{figure}

{In \emph{Gaia} DR2 we identify} a possible companion to J1843-1113 with 2.5$\sigma$ confidence (\citetalias{Ming18}). Our cross-match of J1843-1113's sky position using \emph{Gaia} DR3 returns the same object, \emph{Gaia} 4106823440438736384. Its parallax measurement is $1.06 \pm 0.52$ mas in DR3 and has an S/N of 2.0, nearly double the \emph{Gaia} DR2 S/N. {The FAP in \emph{Gaia} DR3 is similarly on the order of $10^{-2}$, making this another 2.5$\sigma$ association.}

{We combine the \emph{Gaia} DR3 parallax measurement with that from \cite{Desvignes} to yield a distance measurement of $4568^{+3500}_{-1800}$ pc (see \autoref{fig:J1843}). The error on this value has increased by $\sim36\%$ as compared to the DR2 combined distance of $1701^{+3800}_{-105}$ pc. Although the SNR has doubled from DR2 to DR3, the DR3 parallax is much larger than the PTA based measurement, resulting in a broader distribution and larger distance errors. 
}

{While there is no known companion to this pulsar, but we calculate the \emph{Gaia} object's temperature for completeness.} The object has $G_\mathrm{BP}=20.85$, $G_\mathrm{RP}=18.89$ and $C_\mathrm{XP} = 2.0$, so we use the synthetic catalog and estimate $T_\mathrm{eff} \sim3350K$. 

Given the \emph{Gaia} object's {high FAP}, we {emphasize that this is a weak association. J1843-1113 has been studied for decades as a part of PTA experiments with no evidence in timing data to indicate a companion object. However, \cite{binary_comp} find that any pulsar with an unconstrained second frequency derivative, such sas J1843-1113, may host an undetected binary companion with $P_b > 1$~kyr. While unlikely, this may nevertheless be a plausible binary orbit.}

\subsection{J0437-4715: Strong Association} \label{subsec:J0437}

\begin{figure}
\centering
 \vspace{-1\baselineskip}
\includegraphics{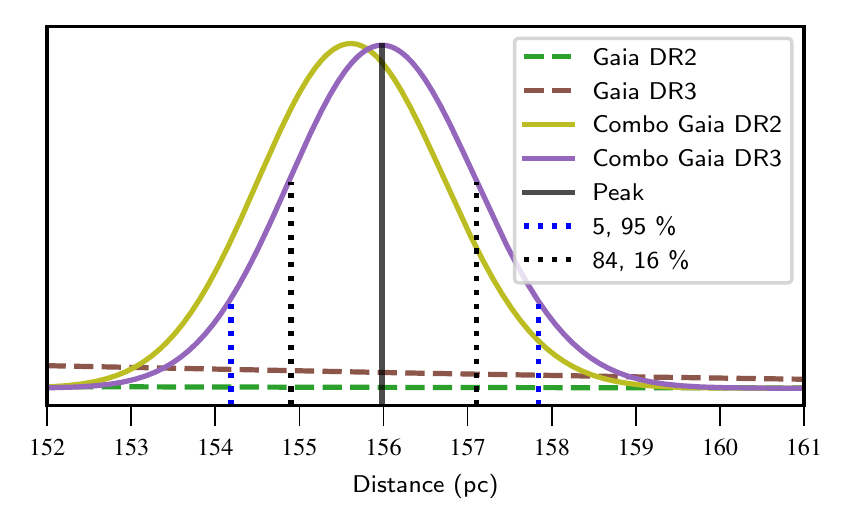}
 \vspace{-1\baselineskip}

\caption{Distances to the system of J0437-4715. The combined curves include parallax based distances from \protect\cite{Dller08}, \protect\cite{Rear16}, and \cite{NG15} in addition to the specified \emph{Gaia} distance. {The distances are not significantly impacted by the \emph{Gaia} parallaxes since the curves are dominated by the ultra-precise VLBI \citep{Dller08} and timing \citep{Rear16} parallaxes.}}
\label{fig:J0437}
\end{figure}

J0437-4715 is in a 5.7 day orbit with a white dwarf companion, and is one of the closest binary MSPs {to Earth} \citep{Johnston, Verbiest}. {\cite{Dller08} report the system's parallax as 6.396 $\pm$ 0.054 mas based on VLBI, and \cite{Rear16} similarly report 6.37 $\pm$ 0.09 mas based on timing parallax. We also incorporate the recent timing complementary parallax measurement from \cite{NG15} of $9.70 \pm 1.11$ mas.} \cite{Rear16} also report a {kinematic} distance measurement of 156.79 $\pm$ 0.25 pc based on the pulsar's well measured orbital period derivative (thus enabling a distance estimate via the Shklovskii effect; \citealt{Shk70}).

We first identified the binary companion to J0437-4715 as \emph{Gaia} DR2 4789864076732331648 (\citetalias{Ming18}) and find the same object in DR3. In \emph{Gaia} DR3 we find that the parallax to this companion object is $7.10 \pm 0.52$ mas and thus our detection's S/N is $7.10/0.52=13.5$. 
In DR2 (\citetalias{Ming18}), the parallax of the associated object is $8.33 \pm 0.68$ mas, thus DR3 has improved the precision of the \emph{Gaia}-based parallax by $\sim 25\%$.

We combine the parallax measurements from \emph{Gaia} DR3, \cite{Dller08}, \cite{Rear16}, and \cite{NG15}. This results in a final distance measurement of $156 _{-1.1}^{+1.1}$~pc. The 5th and 95th percentiles of the distance are $156_{-1.8}^{+1.9}$ pc (see \autoref{fig:J0437}). Our final combined distance varies very little from the \cite{Dller08} and \cite{Rear16} measurements. This is largely due to errors in the \emph{Gaia} parallax measurement being much larger than in other sources, thus having little effect on the combined distance measurement. This is also the case in our analysis of \emph{Gaia} DR2, where we find a distance of $156_{-1.1}^{+1.1}$ pc (\citetalias{Ming18}). 

There is a well-known white dwarf companion to J0437-4715~e.g. \cite{Durant}. Here we find $G_\mathrm{BP}=20.73$ and $G_\mathrm{RP}=19.57$ and use \autoref{eq:temp} to find $T_\mathrm{eff} = 5020 \pm {750}$ K. \cite{Durant} report $T_\mathrm{eff} = 3950 \pm 150$ K for the white dwarf companion to J0437-4715, ${1.4\sigma}$ from our value. 

We compute the FAP of this association with \emph{Gaia} {DR2 and DR3}. In both analyses, we found no positive results in $> 10^{7}$ trials. This indicates a statistically strong, $\gg$5$\sigma$ detection. We are therefore very confident that this companion is indeed the known white dwarf companion to J0437-4715.

\subsection{J1012+5307: Strong Association} \label{subsec:J1012}

\begin{figure*}
\centering
 \subfigure[Right Ascension of J012+5307's \emph{Gaia} companion]{\includegraphics[width=.45\linewidth]{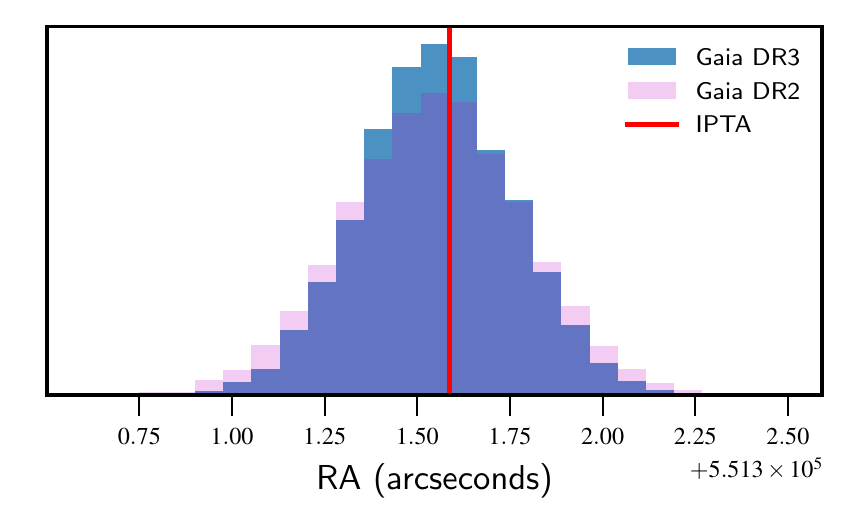}}
\subfigure[Declination of J012+5307's \emph{Gaia} companion]{\includegraphics[width=.45\linewidth]{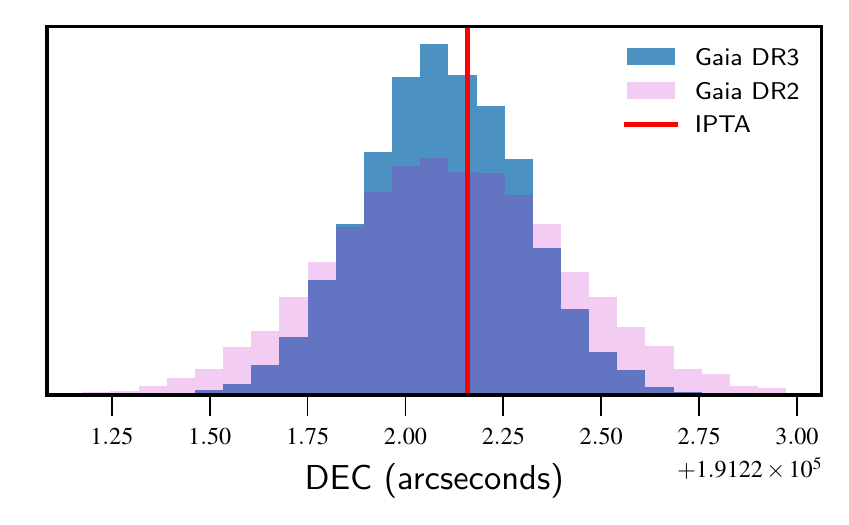}}
\caption {The sky position of the \emph{Gaia} companion to J1012+5307 in \emph{Gaia} DR2 and DR3 for comparison. The pulsar's position in IPTA DR2 is shown in red. The \emph{Gaia} right ascension error decreased by 15\% from DR2 to DR3, and the declination error by $33\%$.}
 \label{fig:J1012}
\end{figure*}

J1012+5307 is in a binary orbit with a low mass white dwarf companion \citep{Nicastro}. \cite{Desvignes} report the parallax of the pulsar as $0.71 \pm 0.17$ mas, and in \cite{Ding} the VLBI-based parallax was found to be 1.17 $\pm$ 0.02 mas. {\cite{Ding23} similarly find the timing-based parallax to be $1.17 \pm 0.05$ mas.}

{In \emph{Gaia} DR2 we identify this binary companion as \emph{Gaia} DR2 851610861391010944 with S/N $= 3.2$ (see \autoref{fig:J1012}; \citetalias{Ming18})}. In \emph{Gaia} DR3, we again identify this object as the companion to J1012+5307. The parallax measurement to this object is $1.74 \pm 0.29$ mas with S/N $= 6.0$. We combine this with the \cite{Desvignes}, \cite{Ding}, {\cite{Ding23}, and \cite{NG15}} measurements {and find a combined parallax measurement of $1.17 \pm 0.02$ mas. This yields a final distance of $845^{+14}_{-14}$ pc. This is $845^{+24}_{-22}$ pc when we use the 5th and 95th percentiles. 
When we calculate the combined distance with the \emph{Gaia} DR2 parallax, the distance is $841^{+10}_{-12}$ pc. This distance is marginally more precise than the \emph{Gaia} DR3 combined distance despite a lower S/N since the DR2 parallax is closer to those reported in \cite{Desvignes}, \cite{Ding},{ and \cite{Ding23}.}}

{ With \emph{Gaia} DR2 we find a FAP $< 7 \times 10^{-8}$, indicating a $> 5 \sigma$ detection. With DR3, we find a FAP of 0 in more than $ 10^{7}$ tests, indicating a $> 5  \sigma$ detection once again. } Furthermore, in DR3 this \emph{Gaia} object has $G_\mathrm{BP}=19.71$ and $G_\mathrm{RP}=19.45$, thus using \autoref{eq:temp} we find $T_\mathrm{eff} = 7407 \pm {1316}$ K. 
{\cite{Sanchez} study this companion via spectroscopic fits (rather than SED fits) and employ correction functions to properly model the extremely low mass companion. They report a temperature of  ${8362^{+25}_{-23}}$ K, --- ${0.8\sigma}$ from what we calculate.} Therefore it is likely that Gaia DR3 851610861391010944 is indeed the previously identified white dwarf binary companion.
 
\subsection{J1024-0719: Strong Association} \label{subsec: J1024}

J1024-0719 is in an ultra-wide binary system ($P_{b} > 200$ years) with a low mass, main sequence star \citep{Bassa,Kaplan}. The spectral type of the companion object was analyzed in \cite{Kaplan} and \cite{Bassa}. 

\begin{figure*}
\centering
\subfigure[Distances to J1012+5307]{\includegraphics[scale=.9]{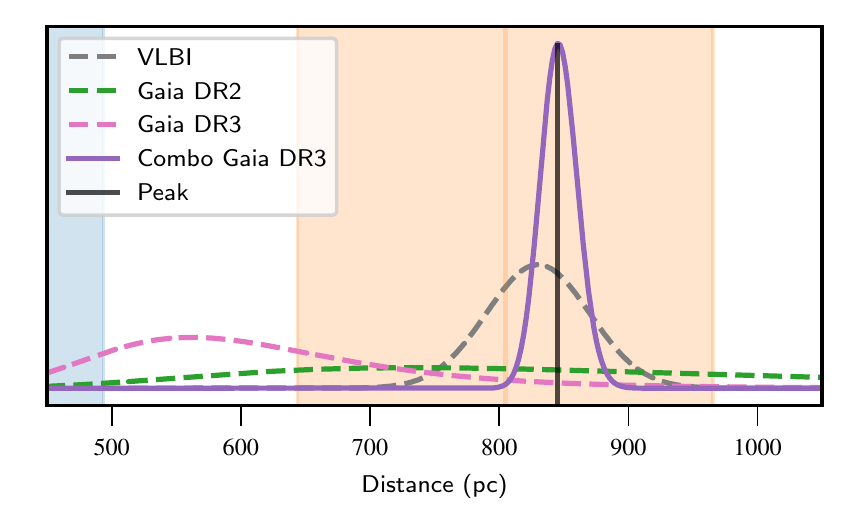}}
 \subfigure[Combined distances to J1012+5307]{\includegraphics[scale=.9]{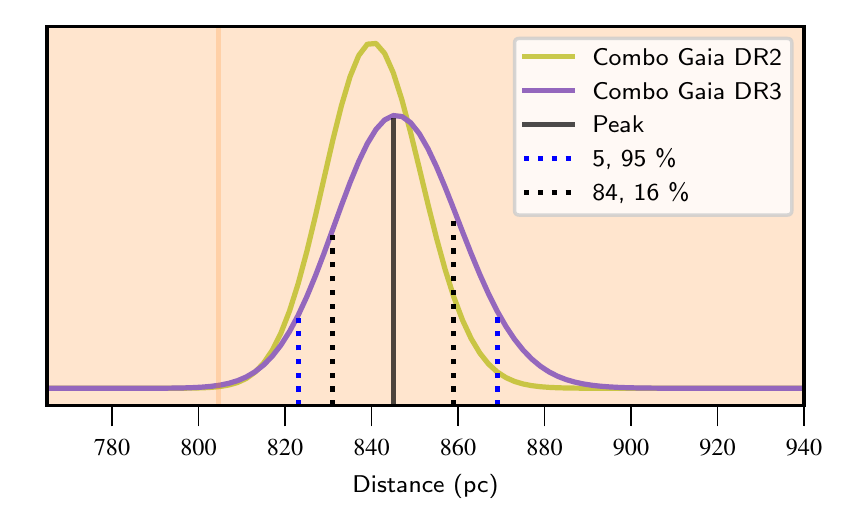}}
\subfigure[Distances to J1024-0719]{\includegraphics[scale=.9]{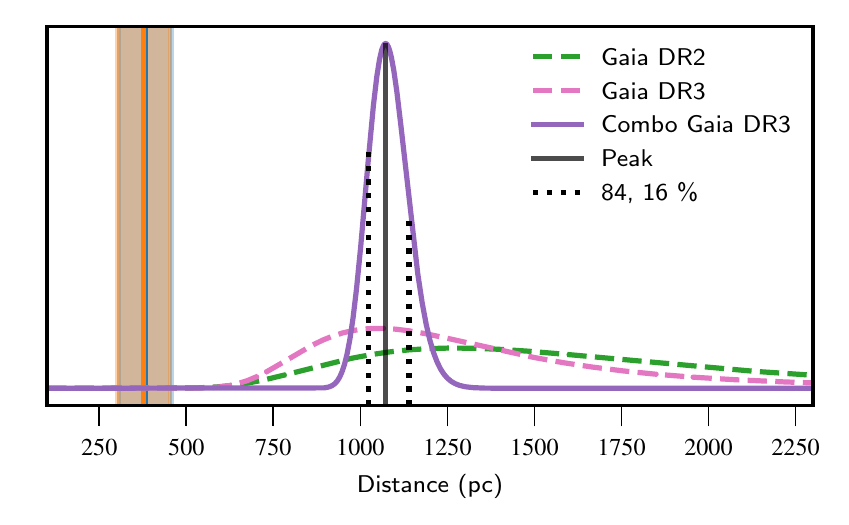}}
\subfigure[Combined distances to J1024-0719]{\includegraphics[scale=.9]{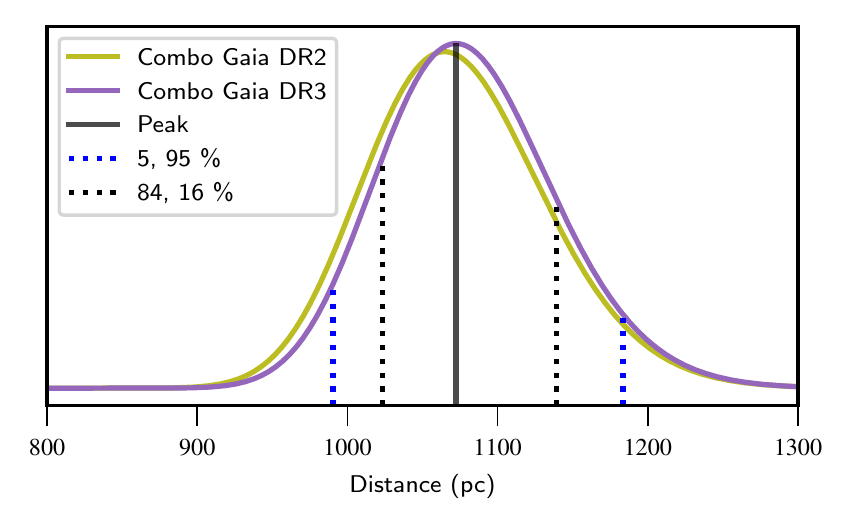}}
 \caption {Distances to J1012+5307 and J1024-0719. All DM distances are shown for comparison only and are not used in any distance calculations. The uncertainties in DM distances assume a standard error of $\pm20\%$. In panels (a) and (b) the combined measurements to J1012+5307 include the \protect\cite{Desvignes} {\cite{Ding23}, and \cite{NG15} parallax measurements} as well as the VLBI distance found in \protect\cite{Ding}. {The combined \emph{Gaia} DR2 distance is more precise than the combined DR3 distance since the DR3 parallax is very small and therefore results in a broader distribution, even when combined with the other parallaxes.
 In panels (c) and (d) the combined measurements include the \citetalias{IPTA}, the NANOGrav 15-yr data~\citep{NG15}, and \cite{Ding23} parallax measurements. All parallaxes correspond to distances more than double the DM distances, which may indicate that the \citetalias{NE2001} and \citetalias{YMW16} underestimate their errors. }}
 \label{fig:J1012_1024}
\end{figure*}

We identify this companion as \emph{Gaia} DR2 and DR3 3775277872387310208 (\citetalias{Ming18}; see also \citealt{Ant}). {In DR2 we find a parallax measurement of $0.53 \pm 0.43$ (S/N $= 1.23$) and in DR3 of $0.86 \pm 0.28$ mas (S/N $= 3.08$).} We combine the DR3 measurement with the pulsar timing based parallax measurements from \citetalias{IPTA}, { \cite{Ding23}, and \cite{NG15}} to yield a new distance measurement of $1064^{+67}_{-49}$ pc. This distance is consistent with the \emph{Gaia} DR2 combined distance of $1155^{+69}_{-50}$ pc (see \autoref{fig:J1012_1024}). 

{The false alarm probability of this identification is $< 10^{-8}$ when using \emph{Gaia} DR2 or DR3, indicating a statistically strong $\gg 5\sigma$ detection in both cases.} Since the object has $C_\mathrm{XP} = 1.8$ in DR3, we must compare the object to the synthetic catalog to estimate its temperature instead of using \autoref{eq:temp}. We find that this object has $G_\mathrm{BP}=20.08$ and $G_\mathrm{RP}=18.26$, data characteristic of a $\sim3500$K star. {\cite{Kaplan} and \cite{Bassa} report the temperature as $3900^{+60}_{-40}$ K and $4050 \pm 50$ K respectively. They classify the object as a cool main sequence star which is consistent with our calculated temperature.}

\section{Discussion and Summary} \label{sec:discussion}

\begin{table*}
 \vspace{-1\baselineskip}

    \centering
    \begin{tabular}{|p{1.9cm}|p{1.4cm}|p{1.6cm}|p{2.0cm}|p{2.4cm}|p{1.5cm}|}
    \hline
    \hline
        Pulsar &  DR2 Detection & DR3 Detection & $T_{\mathrm{eff}}$ (K) & Previous $T_{\mathrm{eff}}$ (K) & Reference\\
        \hline
        \hline
        J0437-4715 & $> 5\sigma$ &$> 5\sigma$ & $5020 \pm {750}$ & $3950 \pm 150$ & \protect{\citetalias{Durant}} \\
        J1012+5307 & $> 5\sigma$ & ${> 5\sigma}$ &$7407 \pm {1316}$& $8362^{+25}_{-23}$ & \protect{\citetalias{Sanchez}}\\
        J1024-0719 & $> 5\sigma$ & $> 5\sigma$ & $\sim3500$ &$4050 \pm 50$ & \protect{\citetalias{Bassa}}\\
        $^\dagger$J1732-5049 & $2.8\sigma$ & ${2.9 \sigma}$& $\sim 4000$ & --&--\\
        $^\dagger$J1747-4036 & N/A & $ {2.3\sigma}$ & $4927 \pm {788}$&--&--\\
        $^\dagger$J1843-1113 & $2.5 \sigma$ &${2.5\sigma}$ &$\sim3350$&--&--\\
         $^\dagger$J1910+1256 & $2.4 \sigma$ & N/A & N/A & -- &-- \\
        \textsuperscript{*}J1949+3106& $3.0 \sigma$ &  N/A & N/A& -- &-- \\
        \hline
        \hline
    \end{tabular}
    \caption{{The strength of identification for each binary system based on FAP and the temperature we calculate for the \emph{Gaia} object. Only associations stronger than ${3.0}\sigma$ (FAPs <$3.19 \times 10^{-3}$) are considered companion detections {and weak associations are denoted by $^\dagger$. False associations are denoted by a `*'.} For J1747-4036, the temperature displayed refers to Object A. Companion temperature information from the specified reference is also shown for comparison. When no previous temperature information is available this is indicated with `--'. }}
    \label{tab:FAPs}
     \vspace{-1\baselineskip}

\end{table*}

We have confirmed associations {of various credibility} to five MSPs in \emph{Gaia} DR3 as well as identified two new candidates for association with J1747-4036, for a total of six possible matches. We also identified a false positive association with J1949+3106 in \emph{Gaia} DR2, from which can conclude that detections must have $> {3.0}\sigma$ confidence to be considered concrete identifications ({see \autoref{tab:FAPs}}). As a complementary means of verification, we calculated companion temperatures, {which can improve detection confidence, see e.g. J0437-4715.}

\begin{table}
    \centering
    \begin{tabular}{|p{1.75cm}|p{2.7cm}|p{2.7cm}|}
    \hline
    \hline
        Pulsar &  DR2 Distance (pc) & DR3 Distance (pc)\\
        \hline
        \hline
        J0437-4715 & $156_{-1.1}^{+1.1}$ &$156 _{-1.1}^{+1.1}$ \\
        J1012+5307 & $841^{+10}_{-12}$ & $845^{+14}_{-14}$ \\
        J1024-0719 & $1155^{+69}_{-50}$ & $1064^{+67}_{-49}$ \\
        $^\dagger$J1732-4036 & $3980 \ (4800, 11000)$ & $3874^{+4100}_{-1400}$ \\
        $^\dagger$J1843-1113 & $1701_{-105}^{+3800}$&$4568_{-1800}^{+3500}$\\
        \hline
        \hline
    \end{tabular}
    \caption{{Comparison of combined distances calculated using the specified \emph{Gaia} parallax measurements. Error bars are the 16th and 84th percentiles of the distance measurements which are in brackets when the 16th percentile is greater than the most probable (peak) distance. Weak \emph{Gaia} associations in need of further verification are denoted by $^\dagger$.}}
    \label{tab:Dist_comp}
     \vspace{-1\baselineskip}
\end{table}

{Looking to the future, parallax errors scale with $T^{-1/2}$, where T is observation time {\citep{Jennings}}. Since the DR3 release comes one year after DR2 which was based on 22 months of data, we expect that DR3 parallax errors will be a factor of $(34/22)^{-1/2} \sim 0.6$ smaller.} The average increase in S/N from DR2 to DR3 is ${53\%}$. The only object for which the S/N did not improve is the companion to J1732-5049, which is in both cases a negative parallax. When this is excluded, the increase
in S/N is ${74}\%$.
Furthermore, the \emph{Gaia} mission has been extended to 2025, promising a DR4 based on 66 months of information\footnote{https://www.cosmos.esa.int/web/gaia/release}. With an additional 32 months of observations, we therefore expect to see a further improvement in parallax of $(66/34)^{-0.5} \sim 0.7$. 

Overall, we find that DM-based distances have underestimated errors. Specifically, the combined distances to J1024-0719, J1732-5049, and J1747-4036 are outside of the assumed DM-based error region of $\pm 20\%$. For J1024-0719 in particular both \citetalias{NE2001} and \citetalias{YMW16} DM models are less than half as large as the parallax distances. These models may have therefore underestimated the electron density in the localized regions of these pulsar systems. 

{Compared to the distances from \emph{Gaia} DR2, the combined distance measurements from DR3 are on average 29\% more precise for systems with previously known companions (i.e. excluding the likely false match with J1843-1113; see \autoref{tab:Dist_comp}). The distances to J0437-4715, J1012+5307, and J1024-0719 are essentially unchanged due to previous parallax measurements which dominate their combined distances. The combined distance to J1732-4026 improved by $30 \%$ as expected, since the \emph{Gaia} measurements are the only known parallaxes. }


We also use the \emph{Gaia} DR3 photometric data to verify our matches. We found that only three \emph{Gaia} companions are suitable to use with \autoref{eq:temp} ($C_\mathrm{XP} < 1.5$): these are J0437-4715, J1012+5307, and J1747-4036. The latter has no published $T_\mathrm{eff}$ value, so we focus our attention on comparing the $T_\mathrm{eff}$ values of J0437-4715 and J1012+5307 to those in the literature. 
Companions to J0437-4715 and J1012+5307 were identified with $\gg5\sigma$ confidence. For J0437-4715's companion, \cite{Durant} report a temperature ${1.4} \sigma$ from what we calculate and {identify the object as a white dwarf. The value we calculate is consistent with this classification.} For J1012+5307, \cite{Sanchez} report a temperature ${0.8\sigma}$ from our calculated value {and identify the companion as a (cool) white dwarf, consistent with the temperature we calculate (see \autoref{tab:FAPs})}. {This method is therefore good enough to determine if the \emph{Gaia} object is within the range of temperatures for a given classification, and results in calculated temperatures within $1.5\sigma$ of published values.}


For the remaining two systems we compare the \emph{Gaia} photometric data to the synthetic catalog in \cite{Jordi} to estimate temperatures.
{We find that the estimated temperatures are useful for determining if the \emph{Gaia} object's photometric data are consistent with expectations for a given classification, although the estimated temperature can only be compared to a previous classification in one case.} 
The temperature of the companion to J1024-0719 {is consistent with} a main sequence object, as is its location in \autoref{fig:hrd}. {This object's photometric data are thus consistent with the classifications made in \cite{Kaplan} and \cite{Bassa}.} 

\emph{Gaia} DR4 will report data based on twice as many transits as DR3. Since \cite{Jordi} reports that error on photometric color data scales as $C_\mathrm{XP} \sim 1/\sqrt{N}$ where N is the number of transits, this photometric data is expected to improve by a factor of $1/\sqrt{2} = 0.7$. Indeed both methods of temperature measurement may become more useful with data from future \emph{Gaia} releases as photometric measurements improve.
{Given this and the expected parallax improvements in \emph{Gaia} DR4 \citep{Fabricius}, we anticipate subsequent improvements in both distance measurements and our verification methods. We will revisit these results again upon the publication of this data release.}

\section{Acknowledgements} \label{acknowledgments}

We thank \cite{Astropy}, \cite{Scipy}, \cite{Numpy}, and \cite{Pygedm} for free Python software which was used to conduct our cross-match and analyze the results. We also thank K. Breivik, N. Khusid, M. Jones, and A. Casey-Clyde for useful conversations. This work has made use of data from the European Space Agency (ESA) mission Gaia (https://www.cosmos.esa.int/gaia), processed by the Gaia Data Processing and Analysis Consortium (DPAC, https://www.cosmos.esa.int/web/gaia/dpac/ consortium). Funding for the DPAC has been provided by national institutions, in particular the institutions participating in the Gaia Multilateral Agreement.

AM thanks the Center for Computational Astrophysics's intern program. Her time with this program has benefited this work. The Center for Computational Astrophysics is a division of the Flatiron Institute in New York City, which is supported by the Simons Foundation. Funding for this project has been provided by the University of Connecticut's Summer Undergraduate Research Fund (SURF). This research was supported in part by the National Science Foundation under Grants No. NSF PHY-1748958, PHY-2020265, and AST-2106552.

\label{sec:references}

\label{lastpage}
\end{document}